\documentclass[letterpaper]{aa}
\usepackage{graphicx}
\begin{document}
\title{\bf A search for very young Planetary Nebulae}
\author{      G. Umana \inst{1}
          \and L. Cerrigone \inst{2}
         \and C. Trigilio \inst{1}
\and R. A. Zappal\`a \inst{2}
         }
\institute{Istituto di Radioastronomia del C.N.R., Stazione VLBI di Noto,
           C.P. 141 Noto, Italy
\and
           Dipartimento di Fisica e Astronomia, Universit\`a di Catania,
           Via S. Sofia 78,
           95123 Catania, Italy}
\offprints{G. Umana, g.umana@ira.cnr.it}
\mail{g.umana@ira.cnr.it}
\date{Received \hbox to1in{\hrulefill}}
\abstract{
Despite  numerous efforts, the transition from Asymptotic Giant Branch (AGB) stars
to Planetary Nebulae (PN) is a poorly understood phase of stellar evolution.
We have therefore carried out interferometric (VLA) radio observations of a
sample of hot post-AGB stars, selected on the basis of their optical and infrared properties.
Ten sources  out of the 16 observed  were detected. This indicates that most of our targets
are surrounded by    a nebula where the ionization has already started.
This definitively determines the evolutionary status of the selected sources and  provides
us with  a unique sample of very
young Planetary Nebulae (yPNe).
A comparison with  another  sample of yPNe confirms our working hypothesis that our targets are
indeed very young, probably
just in the transition toward PN.
Finding transition objects is extremely important as they can provide unique clues for a
better understanding of  this important phase of stellar evolution.
 \keywords{
Stars:AGB and post-AGB -- Planetary Nebulae: general-- radio continuum: stars.}}
\authorrunning{Umana et al.}
\maketitle
%
\section{Introduction}
The formation and  early evolution of Planetary Nebulae (PNe) is a quite obscure phase
of stellar evolution.
Recently HST has provided high-quality images of PNe which quite often reveal jets or bipolar
structures whose origin it is difficult
to explain.
In particular, it is quite challenging to understand how the  spherically symmetric Asymptotic Giant Branch  (AGB)
circumstellar shells transform
themselves into the
extremely asymmetric envelopes observed in evolved PNe.\\
Recent models  propose different agents in shaping PNe such as high-speed, collimated outflows or
 jets that operate during the early
post-AGB evolutionary phase
(Sahai \& Trauger, \cite{Sahai98};  Lee \& Sahai, \cite{Lee}).\\
Very recently, high sensitivity, high resolution MERLIN observations
of IRAS 18276-1431 in the OH maser lines 1612 and 1667 MHz have made it possible,
for the first time,
to detect and measure the circumstellar magnetic field in a PPN
(Bains et al., \cite{Bains}).
 This result suggests that the magnetic field   may play a fundamental
role in the process that shapes the circumstellar envelopes in the post-AGB
evolutionary phase.

During the last few years many observational programs have been devoted to recognizing new
planetary and proto-planetary nebulae
among unidentified IRAS
sources with far infrared colors similar to those of known planetary nebulae
(Pottasch et al., \cite{Pottasch88}; van de Steene \& Pottasch, \cite{vandeS93};
Garcia-Lario et al., \cite{Garcia97a}).
The final aim of these studies is to fully understand the process of formation of Planetary
 Nebulae through
the discovery and analysis of new transition objects in the  short phase between the end
of the AGB  and the onset of the ionization in the nebula.

\noindent The number of known transition objects is extremely small,  tens compared to more than a
thousand PNe
identified in our Galaxy (Kwok, \cite{Kwok94}). This is related to the short
evolutionary timescales involved (a few thousand years).

As post AGB stars/proto-PNe  evolve into early stages of Planetary Nebulae, part
of the shell begins to be ionized by
the central star.
The post-AGB objects detected  by IRAS extended from non variable OH-IR stars to M, K, G, F, A supergiant types.
This sequence appears to represent the evolution of the post-AGB stars towards hotter spectral types and
we should also expect to find some B-type stars among the hottest post-AGB stars.
There is a small sub-group of B-type stars, called BQ[~] stars, defined as
$B_\mathrm{e}$ with forbidden emission lines, and Parthasarathy \& Pottasch (\cite{Part89}),
on the basis of their IR excess,  recognized them as potential candidates to be new
transition objects.

BQ[~] stars, however, are not a well defined group, and there is still a controversy on their
evolutionary stage. Parthasarathy \& Pottasch (\cite{Part89}) suggested that
at least a fraction of them  could be hot post-AGB stars in an early stage of their evolution
as proto-PNe.
\noindent
The common characteristics of these kinds of star are:
\begin{itemize}
\item numerous permitted and forbidden emission lines of several elements, in addition
 to the absorption
lines typical of a B1 I-II star;
\item underabundance of carbon and metals;
\item  high galactic latitude;
\item  presence of PN-type detached cold circumstellar shells.
\end{itemize}
\noindent These characteristics are typical for low-mass, highly evolved post-AGB stars.

\noindent In some cases the central star may just have become hot enough to photoionize
the circumstellar
envelope ejected during the previous AGB phase,  as
the presence of low excitation nebular emission lines in the spectrum seems to indicate.

\section{Observations and Results}
\subsection{Sample selection}
In order to find new very young Planetary Nebulae we have selected  a sample
of 16 hot post-AGB stars from the most recent compilations, namely
Parthasarathy, (\cite{Part93a}); Conlon et al., (\cite{Conlon}) and  Parthasarathy et al.,
(\cite{Part00a}).
All the selected  candidates  have high galactic latitude,
infrared excess, spectral type B1 I-II and  nebular emission lines in their
spectrum (Table~\ref{stars}).
In particular the last two  requirements maximize the possibility of detecting  a radio nebula.
Moreover in one of the proposed targets, namely
SAO 85766,  the strong variations observed in the spectrum
 suggest that it has rapidly evolved in the last 50 years
(Parthasarathy et al., \cite{Part00b}) and its  evolutionary state appears to be similar
to that observed in the case of the
hot post-AGB star SAO 244567,
where a radio nebula has been detected (Umana et al., \cite{Umana2004}).

The detection of radio nebulae associated to the selected targets
would definitively establish  the evolutionary status of this kind of objects, producing
a unique sample of very young PNe.\\
Successive higher resolution radio observations
of the detected targets will allow  us  to  study nebular morphologies
in the very early phases of  evolution of PNe
and thus clarify at what stage of the AGB-PN evolution those asymmetries are
established , providing  strong clues to the process  shaping PNe.

\subsection{The VLA  data}

The observations were carried out  using the VLA \footnote{The Very Large Array
is a facility of the
National Radio Astronomy Observatory which is operated by Associated Universities,
Inc. under cooperative agreement with the
National Science Foundation}
in two different runs, the first on June 8, 2001, from 06:06 UT to 11:33 UT, and the second
on June 10, 2001 from 18:56 UT to 20:24 UT.\\
We observed  our targets at  8.4~GHz (X-Band), using two independent 50 MHz bands.
The choice of the observing frequency is a
compromise between the sensitivity of the instrument and the
higher probability of  observing in the optically thin region of the spectra.
The array was in CnB configuration, providing a typical
beam size of $\sim 2.5\arcsec$.

A typical observing cycle consisted of 10~min integration time,  preceded and followed
by a 2-min observation of the phase calibrator. This basic sequence
was repeated at least 3 times in order to improve the signal to noise ratio
and thus to ensure the necessary sensitivity for the detection of weak sources.
The flux density scale was determined by observing 3C286 and 3C48, while
phase calibration  was obtained from observations
of phase calibrators close to our targets.

       The data  processing was performed  using the standard programs
 of the NRAO {\bf A}stronomical {\bf I}mage  {\bf P}rocessing  {\bf S}ystem (AIPS).
 To achieve the highest possible signal to noise ratio, the mapping process
 was performed  using the natural weighting and the dirty
 map was {\bf CLEAN}ed down as close as possible to the theoretical noise.
 To estimate the noise level in the maps   we analyzed an area on the map,
 whose dimension is of the order of more than 100 $\theta_\mathrm{syn}^{2}$,
 away from the phase center and free from evident radio sources.
 Its consistency with the expected theoretical noise was also checked.\\
 Of 16 observed sources, we detected a total of ten.
The source position and the flux density determination were obtained by fitting a
 gaussian brightness distribution (JMFIT).
 The uncertainty in the source position is a function of the rms in the map,
 the resolution ($\theta_\mathrm{syn}$), and  the peak flux density of the
 source in the image ($F_\mathrm{peak}$), namely:

\begin{equation}
\sigma_\mathrm{pos} \sim \frac{rms~ \theta_\mathrm{syn}}{2F_\mathrm{peak}}
\sim \frac{\theta_\mathrm{syn} }{2 \frac{S}{N} }
\label{eq1}
\end{equation}

where $\frac{S}{N}$ is the signal to noise ratio for the source detection.
For the worst case in our sample we derive a $\sigma_\mathrm{pos}$ of about
$0 \farcs 12 $.

Our results are summarized in Table~1, where the radio flux density, with
associated error ($\sigma$) and the rms of the map, are reported.
The error associated to the flux density estimation is a combination of
the rms in the map plus the amplitude  calibration error.
Therefore, $\sigma$ is derived from:
\begin{equation}
\sigma = \sqrt{ (rms)^{2} + (\sigma_{\mathrm{cal}} F_\mathrm{3.6cm})^{2}}
\label{eq2}
\end{equation}
where $\sigma_\mathrm{cal}$ is the error associated to the flux calibration,
 which is of the order of 2-3 \%, and $F_\mathrm{3.6cm}$ is the measured flux density.
At CnB resolution all the detected targets  are point-like or slightly resolved
 radio sources, i.e.
$\theta_\mathrm{source} \leq 2\arcsec$. \\

All the sources in our sample were checked for 20~cm emission  using the
NRAO VLA Sky Survey (NVSS, Condon et al., \cite{Condon98}).
Only two of our detected targets were present in the survey, namely \object{IRAS~18442}
and \object{IRAS~19336},
with fluxes at 1.4 GHz of $8.4 \pm 0.6$ and $8.8 \pm 0.6$~mJy
respectively.\\

It is not easy to verify whether the non-detections are related
 to a  larger
distance or due to an  intrinsic characteristic of the source
such as the evolution of the ionization structure.
In fact, distances to planetary nebulae are notoriously  very difficult
to be estimate.

Traditionally,  distances are derived through statistical methods, which are
based on a particular characteristic of the nebula, assumed to be invariant.
For example, the Shklovsky method assumes that all PN have the same ionized  masses
(Cahn \& Kaler \cite{CK71}). Since this  is often not the case,
statistical methods are considered not to be  reliable (Terzian, \cite{Terzian93}).\\
Recently, Gauba et al. (\cite{Gauba03}) modelled the near and far-infrared flux distribution of a
small sample of hot post-AGB candidates, with IRAS colors similar to those of PNe.
Their sample includes four of our targets, namely IRAS 17074+1845, IRAS 17423-1755, IRAS 19157-0247
and IRAS 19399+2312.
Using the DUSTY code (Ivezic et al. \cite{DUSTY}), the authors modelled the circumstellar shells of their sample and derived various physical parameters, including an estimate of the distance.
They derived a distance of 3.7, 3.7, 0.9 and 0.7 Kpc respectively to \object{IRAS~17074+1845}, \object{IRAS~17423-1755}
\object{IRAS~19157-0247} and \object{IRAS~19399+2312}, with an uncertainty  of
about $10\%$ for each source, depending
on the assumed core mass.\\
Their results indicate,  at least for a small fraction of our sample, that
the lack of detectable radio flux should be related to intrinsic factor  and not to the distance.
As a matter of fact, the undetected sources IRAS 19157-0247 and IRAS 19399+2312 have distances much smaller than the detected source IRAS 17423-1755.\\

We stress again the difficulties in deriving distances to PNe, regardless of the used method, as it will lead to uncertainties by a typical factor of 2 or 3.
As an example, the distance to \object{IRAS~17423-1755} has been derived by different authors,  using different methods, to range from 8.3 kpc (Borkowski \& Harrington, \cite{Borkowsky01}) to 5.4 kpc
(Riera et al., \cite{Riera03}).
Therefore, results presented by Gauba et al. (\cite{Gauba03}) provide only an indication
and  should be taken with  caution.

\smallskip

\tiny{
\begin{table*}
\caption[]{Programme Stars}
\label{stars}
~\\
~\\
\begin{tabular}{|cccccc|cc|ccc|} \hline\hline
{\em Name} & IRAS   & $\alpha_\mathrm{J2000}$   & $\delta_\mathrm{J2000}$&  $\alpha_\mathrm{radio}$   & $\delta_\mathrm{radio}$
 &  \em{V}.& \em{S.T.} & {$F_\mathrm{3.6cm}$} & $\sigma$ & rms \\
& &\footnotesize{1} & \footnotesize{1} & &  & &  & mJy &  mJy & mJy\\
\hline
OY~Gem & 06556+1623   & 06 58 30.3  & +16 19 25   & 06 58 30.3  & +16 19 26.2 &  $11.5^{\footnotesize{(2)}}$ & $BQ^{\footnotesize{(2)}}$ & 0.55 & 0.03 & 0.02 \\
 Hen 3-1347  &  17074-1845    & 17 10 24.1  & -18 49 01 &  &  & $9.5^{\footnotesize{(4)}}$ & $B3IIIe^{\footnotesize{(4)}}$ &   & & 0.03 \\
       & 17203-1534    & 17 23 11.8  & -15 37 15 &   &  & $12.5^{\footnotesize{(4)}}$  & $BIIIe^{\footnotesize{(4)}}$ &  &  & 0.03 \\
LS 4331  & 17381-1616    & 17 40 59.9  & -16 17 58   & 17 41 00.1  & -16 18 12   & $12.2^{\footnotesize{(4)}}$ & $B1Ie^{\footnotesize{(4)}}$  & 1.42  &  0.05 & 0.03\\
Hen 3-1475  & 17423-1755   & 17 45 14.2  & -17 56 47   & 17 45 14.1  & -17 56 45  & $11.0^{\footnotesize{(2)}}$ & $Be^{\footnotesize{(2)}}$ & 0.26  &  0.03 & 0.03 \\
SAO 209306  & 17460-3114    & 17 49 16.5  & -31 15 18  & 17 49 16.5  & -31 15 18  & $7.9^{\footnotesize{(4)}}$   & $B0IIe^{\footnotesize{(4)}}$  & 1.29  & 0.05 & 0.04\\
V886 Her  & 18062+2410    &  18 08 20.1    & +24 10 43  & 18 08 20.1  & +24 10 43 & $11.6^{\footnotesize{(4)}}$ & $B1IIIe^{\footnotesize{(4)}}$  & 1.46  &  0.05 &0.03 \\
LS 63  & 18371-3159    & 18 40 21.6  & -31 56 49   & 18 40 22.0  & -31 56 49 & $11.9^{\footnotesize{(4)}}$ & $B1Ie^{\footnotesize{(4)}}$ & 0.62  & 0.03 & 0.03 \\
LS 5112 & 18379-1707   & 18 40 48.5  & -17 04 36   &  &  & $11.8^{\footnotesize{(4)}}$ & $B1IIIe^{\footnotesize{(4)}}$  &  &  & 0.03 \\
       & 18435-0052    & 18 46 06.9  & -00 48 55   &  &  & $11.0^{\footnotesize{(4)}}$ & $B2II^{\footnotesize{(4)}}$   & & &  0.03\\
BD -11 4747 & 18442-1144  & 18 47 04.8  & -11 41 02    & 18 47 04.0  & -11 41 12  & $9.3^{\footnotesize{(2)}}$ & $Be^{\footnotesize{(2)}}$ & 19.21 & 0.60 & 0.05 \\
LS IV -02 29 & 19157-0247  & 19 18 22.5  & -02 42 09   &  &   & $11.5^{\footnotesize{(4)}}$ & $B1III^{\footnotesize{(4)}}$ & &  &0.03\\
       & 19336-0400    & 19 36 17.5  & -03 53 24     & 19 36 17.5  & -03 53 25   & $12.5^{\footnotesize{(4)}}$ & $B1Ie^{\footnotesize{(4)}}$ & 9.74  & 0.29 & 0.03\\
LS II +23 17 & 19399+2312  & 19 42 05.5  & +23 18 59  &  &   & $10.4^{\footnotesize{(4)}}$  & $B1III^{\footnotesize{(4)}}$ & & &0.03\\
LS IV -12 111 & 19590-1249 & 20 01 49.8  & -12 41 17 &  20 01 49.8  & -12 41 17 & $11.3^{\footnotesize{(4)}}$ & $B1Ie^{\footnotesize{(4)}}$& 2.76  & 0.08 & 0.03 \\
LS II +34 26& 20462+3416  & 20 48 16.6  & +34 27 24    & 20 48 16.6  & +34 27 24  & $11.1^{\footnotesize{(3)}}$ & $BIe^{\footnotesize{(3)}}$ & 0.42  &  0.03 & 0.03\\
\hline \hline
 \end{tabular}
~\\
~\\
\footnotesize{1) Optical coordinates are FK5 2000.0 as reported by SIMBAD database.}\\
\footnotesize{2) Parthasarathy, \cite{Part93a}}\\
\footnotesize{3) Parthasarathy, \cite{Part93b}}\\
\footnotesize{4) Parthasarathy et al., \cite{Part00a}}\\
\end{table*}
}
\normalsize

\section{Notes on individual sources}

To point out any characteristics common to the detected sources, besides selection criteria,
we summarize, in the following, the physical properties of each source as derived from the literature.
 When available, a previous radio measurement is reported and compared with our result.
All the targets in our sample are associated witn an IRAS source. When a discrepancy between radio,
optical and IRAS coordinates is present, a comment is added at the end of each source section.\\

\noindent
{\bf \object{IRAS 06556+1623}}

When it was discovered, in 1932, this star showed some P Cygni line profiles in its spectrum.
Ciatti et al. (\cite{Ciatti}) pointed out
the presence, in the spectrum, of several strong   emission lines.
The continuum emission was well represented by a B0.5 star and  the line ratios
for [\ion{O}{i}], [\ion{O}{ii}], [\ion{O}{iii}]
were not compatible with a unique set of density and temperature of the nebula.
This suggests that
the star is  surrounded by an extended gaseous shell composed of regions of different temperatures,
densities and velocities (Klutz  \& Swings, \cite{Klutz}).\\
By comparing their spectrophotometric data with other data previously reported by other authors,
Arkhipova \& Ipatov (\cite{Arkhipova82}) concluded that the volume and density
of the gaseous shell had been increasing for the last 40 years.
The more recent spectra are those obtained by Jaschek et al. (\cite{Jaschek}), which were
collected between 1990 and 1993.
Many lines are present, including several classic nebular lines such as [\ion{O}{iii}].
No significant differences with previous similar work are reported.
Attempts to detect radio emission were made  by Sistla \& Seung (\cite{Sistla}) at 8.4 GHz and
by Altenhoff et al. (\cite{Altenhoff}) at 10.7 GHz, but both studies the sensitivity was not high
enoigh to detect the source, the rms being of the order of 5~mJy.

\noindent
{\bf \object{IRAS~17381-1616}}

The star was included in a  sample of  PN candidates  selected  from the IRAS Point Source Catalog
on the basis of
their far-IR colours (Preite-Martinez, \cite{Preite}).
\object{IRAS~17381-1616} was included in a list of objects that were probably not PN but PN-related.
Parthasarathy et al. (\cite{Part00a})
classified it as a
B1Ibe star, with H$\beta$ and H$\gamma$ emission lines, suggesting that the object was a post-AGB star.

The interferometer was pointed at \object{LS~4331}, which is believed to be the optical
counterpart of \object{IRAS~17381-1616}. However, we did not find the radio source
at the  expected optical coordinates, but  we detected a $1.42 \pm 0.05$~mJy radio
source at  a radio position of $\alpha=17:41:00.05$ and $\delta=-16:18:12.45$,
corresponding to a shift of  $2.16\arcsec$ in right ascension and  about $16\arcsec$ in declination, or
about 6.5 $\theta_\mathrm{syn}$.

However the radio positions are shifted by only  $\Delta \alpha = 2\farcs 16 $ and
$\Delta \delta = 1\farcs 15$ with respect to the
IRAS coordinates, well inside the IRAS ellipse error ($34\arcsec \times 8\arcsec$ ).
We can thus conclude that our radio source is associated to the IRAS source.

\noindent
{ \bf \object{IRAS~17423-1755}}

The source was believed to be a massive population I B-type star,
but Bobrowsky et al.  (\cite{Bobrowsky95}) recognized it as a new planetary nebula.

High resolution spectroscopy and optical images reveal
a complex  bipolar morphology: a dense central region (disk) from which
large-scale flows collimate into bipolar jets
(Bobrowsky et al. \cite{Bobrowsky95}; Riera et al., \cite{Riera};
Borkowski \& Harrington, \cite{Borkowsky01}).

Parthasarathy et al. (\cite{Part00a})
found Balmer lines having a  P Cygni profile, blue-shifted by $-400\,\mathrm{km\,s}^{-1}$
and from the analysis of H$_{\alpha}$ profiles
Sanchez Contreras \& Sahai (\cite{Sanchez}) pointed out
the presence of two different winds, one of these probably associated to the post-AGB
outflow prior to its interaction with the AGB wind.

Knapp et al. (\cite{Knapp}) detected it at 8.4 GHz with the
Very Large Array (NRAO) and found the flux to be $0.30\pm 0.04$~mJy,
with an excitation parameter of $1.8$, compatible with a B3 spectral type.
Our flux determination is in agreement, within the errors, with previous results.

\noindent
{\bf \object{IRAS~17460-3114}}

Crampton (\cite{Crampton}) classified it as an O7.5 star in a \ion{H}{ii} region; this was
confirmed by Dachs et al. (\cite{Dachs}). Gies (\cite{Gies}) and
Mason et al. (\cite{Mason}) found a O8V spectral type.
On the basis
of its far-IR colors similar to PNe and its spectral type
Parthasarathy  (\cite{Part93a}) classified \object{IRAS~17460-3114} as a  hot post-AGB star.

Radio measurements were made by Ratag et al. (\cite{Ratag}), who found a flux of $0.8$~mJy at
$4.8$~GHz using the VLA.\\
We detect this source with a flux density at 8.4~GHz of  $1.29\pm0.05$~mJy.
Combining our results with data from Ratag et al. (\cite{Ratag}), we derive
a spectral index $\alpha=0.85$, consistent with an optically thick nebula, at least between 4.8
and 8.4~GHz as we expected from a very young, and thus dense, PN.

\noindent
{\bf \object{IRAS~18062+2410}}

According to the HDE catalogue its spectral type was A5 in 1940, but
Downes \& Keyes (\cite{Downes}) classified it as a Be star.
Arkhipova et al. (\cite{Arkhipova99}) and Partasarathy et al. (\cite{Part00b})
analyzed the high resolution spectrum of this star and found several permitted
and forbidden emission lines: the presence of [\ion{N}{ii}] and [\ion{S}{ii}] lines indicate
the presence of a low excitation nebula.
Rapid photometric variability has also  been  detected (Arkhipova et al., \cite{Arkhipova01a})
which, together the observed spectral variations,
appears to indicate that the star has been  evolving rapidly in the last 150 years.

\noindent
{\bf \object{IRAS~18371-3159}}

Preite-Martinez (\cite{Preite}) selected this star from the IRAS PSC as a planetary nebula candidate on the basis of its far-IR colors.
Parthasarathy et al. (\cite{Part00a}) found its spectral type to be B1Iabe, which combined
with  its high galactic latitude, far-IR colors
similar to PN, and Balmer line emission, makes it  a post-AGB candidate.

\noindent
{\bf \object{IRAS~18442-1144}}

This source is very near to \object{BD-11~4747} and  possibly constitutes  a binary
system or is  identifiable with it. Corporon \& Lagrange (\cite{Corporon}) did not find
any evidence of binariety.
Parthasarathy et al. (\cite{Part00a}) found an A3V spectral type for \object{BD-11~4747},
but they could not observe the optical counterpart of
\object{IRAS~18442-1144} because it was  probably too faint to detect.
On the basis of the presence of [\ion{N}{ii}] lines in the spectrum \object{IRAS~18442-1144}
may be a post-AGB star or a low excitation PN.

We found no source at the coordinates of \object{BD-11~4747}, which is reported as the optical
counterpart of \object{IRAS~18442-1144}.
A  distinct object was detected at $RA=18:47:04$ and $DEC=-11:41:12$, with a
flux density of $19.21\pm0.06$~mJy.

The  radio positions are shifted by only  $\Delta \alpha =3\arcsec$ and
$\Delta \delta = 1\arcsec$ with respect to the
IRAS coordinates, well inside the IRAS ellipse error ($ 30\arcsec \times 7\arcsec$).
We can thus conclude that the  radio source is associated to  \object{IRAS~18442-1144}.\\
If our data are combined  with those  from NVSS it appears that the nebula around
\object{IRAS~18442-1144} is optically thick between 1.4 and 8.4 GHz.

\noindent
{\bf \object{IRAS~19336-0400}}

Downes \& Keyes (\cite{Downes}) found [\ion{N}{ii}] and [\ion{S}{ii}] lines in the  spectrum of \object{IRAS~19336-0400}, suggesting the presence of a nebula.
Parthasarathy et al. (\cite{Part00a}) classified it as a B1 supergiant with Balmer
lines in emission, but without [\ion{O}{iii}] line,
thus suggesting it is a very young and low excitation planetary nebula.
\object{IRAS~19336-0400} was detected by van de Steene \& Pottash (\cite{vandeS95}) with the WRST
at 4.8~GHz as a $10.5 \pm 1.1$~mJy radio source. The source is  also present in the
NVSS with a flux density at 1.4~GHz of $8.8\pm{0.6}$~mJy.
We measured a 8.4~GHz radio flux density of $9.74\pm 0.3$~mJy. \\
Combining our data  with those from the literature we derive a quite flat radio spectrum,
indicating a nebula which is optically thin between 1.4 and 8.4 GHz.

\noindent
{\bf \object{IRAS~19590-1249}}

Ultraviolet, optical and infrared observations of \object{IRAS~19590-1249}
(McCausland et al. \cite{McCausland}; Conlon et al. \cite{Conlon}) suggest that
this B0 supergiant is in the post-AGB evolutionary stage, evolving
into a low excitation planetary nebula.
By comparing the atmospheric parameters of the stars with theoretical post-AGB evolutionary
tracks a central mass of 0.67 $M_{\sun}$
has been estimated (Conlon et al. \cite{Conlon}), implying that further evolution
can occur on a timescale not more than  hundreds of years.
Photometric variability has been reported (Arkhipova et al., \cite{Arkhipova02}), with a pattern
similar to what has been observed  in other two of our detected targets,
namely \object{IRAS~18062+2410} and \object{IRAS~20462+3416}, which appears to be attributable to
their common evolutionary status.\\

\noindent
{\bf \object{IRAS~20462+3416}}

Turner \& Drilling (\cite{Turner}) classified this star as a B1.5 supergiant,
but,
on the basis of its high galactic latitude, far-IR colours similar to PN, dust shell parameters
and flux distribution, Parthasarathy (\cite{Part93b}), concluded that it was a low mass post-AGB star, with a detected cold dust shell.
The optical spectrum of \object{IRAS~20462+3416} consists of the spectrum of a B1.5 supergiant plus
the emission spectrum of a low-excitation nebula, with no [\ion{O}{iii}] lines (Parthasaraty, \cite{Part93b};
Smith \& Lambert, \cite{Smith}).
Strong spectral variations were observed (Smith and Lambert, \cite{Smith};
Garcia-Lario et al., \cite{Garcia97b}),
which are interpreted in terms of mass-loss episodes.
Evidence of mass-loss has been reported also by Arrieta \& Torres-Peimbert (\cite{Arrieta}) from the
analysis of P Cyg profiles observed in both UV and optical spectra.
As in other hot post-AGB star, the star exhibited rapid, irregular light variation, that can be
attributed  to a variable stellar wind and/or to some kind of
stellar pulsations (Arkhipova et al., \cite{Arkhipova01b}).\\

 We want to stress that only  few of our targets have been studied in detail.
In particular, multi-epoch photometric and spectroscopic studies are available only for
\object{IRAS 06556+1623}, \object{IRAS~17423-1755}, \object{IRAS~18062+2410},
\object{IRAS~19590-1249} and
\object{IRAS~20462+3416}. Such studies indicated a  strong variability in both continum
and spectral lines.\\
We may conclude that, besides selection criteria and the observed free-free emission, a possible common characteristic to our sources is  strong spectral variability which may indicate fast evolution through episodic mass-loss events.

\begin{table*}
\caption{Summary of nebular characteristics of detected targets.
Angular diameters are corrected following van Hoof (\cite{vanHoof});
The mean emission measure (EM) has been calculated from
the formula of Terzian \& Dickey (\cite{Terzian});
IRE are derived following  Pottasch (\cite{Pottasch84a})
}
\label{detected}
~\\
~\\
\begin{tabular}{|lccccc|}
\hline \hline
 IRAS ID &  Diameter & $T_\mathrm{B}$ &  $EM$ & IRE & $T_\mathrm{dust}$ \\
 &    [arcsec] & [K] & [$10^{4}\mathrm{cm}^{-6}$pc] &  &[K]  \\
\hline
\emph{06556+1623} &  2.1         & 2.3        & 6.3 & 194 & 181 \\
\emph{17381-1616} &  $\leq 2.0$  & $\geq 8.9$ & $\geq 18.8$ & 31 & 218 \\
\emph{17423-1755} &  $\leq 2.0$  & $\geq 1.6$ & $\geq 3.4$ & 2984 & 130 \\
\emph{17460-3114} &  1.1         & 27         & 56.6 & 248 & 204\\
\emph{18062+2410} &  $\leq 2.0$ & $\geq 9.1$ & $\geq 19.3$ & 106 & 253 \\
\emph{18371-3159} &   $\leq 2.0$ & $\geq 3.9$ & $\geq 8.2$  & 187 & 187\\
\emph{18442-1144} &  1.8         & 148        & 314.8 & 11 & 129 \\
\emph{19336-0400} &  1.5        & 108        & 229.8  & 14 & 187\\
\emph{19590-1249} &  1.9         & 19.1       & 40.6  & 21 & 134 \\
\emph{20462+3416} &  2.2        & 2.2        & 4.6 & 186 & 114 \\
\hline\hline
\end{tabular}
\end{table*}

\section{Nebular Characteristics}
Since the aim of this work is to find new objects in the very initial stage
of PNe evolution, we need to check if our detected objects are indeed
young PNe.
We will first determine the radio  characteristics of our sample
and then we will check how they compare with those of   young PNe
as derived from previous, more extended radio  surveys.

\subsection{ Brightness Temperature}
>From measurements of the radio continuum of  PNe it is possible to determine
a distance-independent parameter, namely  the  brightness temperature
($T_\mathrm{B}$), which is defined as:
\begin{equation}
T_{\mathrm{B}}=(\frac{c^{2}}{2 \pi k \nu^{2}})\frac{F_{\nu}}{\theta^{2}}
\label{eq4}
\end{equation}
where $F_{\nu}$ is the  flux density at frequency $\nu$, and $\theta$ is
the angular diameter of the nebula.
For $\nu=8.4$~GHz Eq.(\ref{eq4}) becomes:
\begin{equation}
T_\mathrm{B}=25(\frac{ \theta}{arcsec})^{-2}\frac{F_\mathrm{8.4~GHz}}{\mathrm{mJy}}  \mathrm{K}
\label{eq5}
\end{equation}
At the resolution of our observations  4 detected targets are point-like.
In these cases  we can derive only a lower limit for the brightness temperature,
assuming for the sources an angular diameter $\leq 2 \arcsec$.
For the rest of the target the angular size has been derived  by a Gaussian fitting
of the source  in the map (see Sect 2.2) and by performing a Gaussian deconvolution
of the fitted source size.

If we indicate with $\theta_\mathrm{syn}$ the FWHM of the point-source response of the interferometer
and with $\theta_\mathrm{G}$ the fitted FWHM size, the resultant deconvolved FWHM size ($\theta$) is:
\begin{equation}
\theta=\sqrt{ (\theta_\mathrm{G})^{2} - (\theta_\mathrm{syn})^{2} }
\end{equation}
which, for resolved sources, can be significatively smaller than the nominal resolution
of the interferometer.

The obtained  angular sizes were then  corrected for the appropriate factor ($\gamma$)
following van Hoof (\cite{vanHoof}),
\begin{equation}
\gamma(\beta)= \frac{ 0.36} {1 + 0.7994 \beta^{2}} + 1.1106
\label{eq6}
\end{equation}
where $\beta$ represents the ratio between {\bf  $\theta_\mathrm{G}$ and  $\theta_\mathrm{syn}$.}\\
The results are summarized in Table~\ref{detected}.
As the nebula expands,  as a consequence of PN evolution, its surface brightness
is expected to decrease. Thus $T_\mathrm{B}$ is usually considered  an age indicator,
as  high $T_\mathrm{B}$ should characterize compact,  very young PNe.
$T_\mathrm{B}$ measured for large samples of PNe are in the range between 10000
and 0.1 K (Kwok, \cite{Kwok90}). Our  results indeed indicate  very low  $T_\mathrm{B}$ values for our sample.
This, however, can be still in agreement with the hypothesis of young age if one assumes that at the
observing frequency the optical depth of the nebula  is very small.\\
 Among the targets of our sample  there are three sources, namely IRAS~17460-3114, IRAS~18442-1144 and
IRAS~19336-0400, whose radio spectrum can be inferred on the basis of data from the literature and the results presented
in this paper (see Sect 3).
It can be argued that the first two sources appear to be optically thick, at least between 20 and 3.6~cm.
Unfortunately, the available spectral information is too poor and a two-point spectrum, in one case
obtained very far apart in frequency, is not sufficient to draw any conclusion on the spectral index.

\subsection{Emission Measure}

Another physical parameter whose high value should characterize a young PN  is the electron density
which is expected to decrease as the nebula ages and expands.\\
The mean emission measure ($<EM>$),  as derived from radio
observations, is an averaged property of the ionized nebula and, therefore, may better
represent the overall density structure of PNe.

If the source is optically thin we can use the observed radio flux density
plus its angular size to derive the mean emission measure as follows  (Terzian \& Dickey, \cite{Terzian}):
\begin{equation}
<EM>=\frac{ \int_{\Omega} EM  d\Omega}{ \Omega} =\frac{ 5.3 \times 10^{5} F_\mathrm{8.4~GHz}}{ \theta^{2} }
\label{eq7}
\end{equation}
where $F_\mathrm{8.4~GHz}$ is the measured radio flux density, in mJy, at 8.4~GHz,
and $\theta$, in arcsec, the angular dimension of the
radio emitting region at that frequency. The mean emission measure is in cm$^{-6}$pc.\\
Young PNe should have emission measures of the order of $10^{6}-10^{8}$cm$^{-6}$pc (Terzian \& Dickey, \cite{Terzian};
Kwok et al., \cite{Kwok81})
and, as reported in
Table~\ref{detected},
our sample, in general, appears to have a more diluted nebula.

\subsection{Infrared Excess}
A young planetary nebula is still surrounded by the remnant of the dusty envelope of its AGB progenitor and emits
most of its energy in the far
infrared as thermal radiation from heated dust grains.
The infrared excess (IRE) is defined as the ratio of the observed total far infrared flux
($F_\mathrm{IR}$) over the expected total infrared flux.
On the hypothesis that the far infrared flux is due to thermal emission from dusty grains heated by
Ly$\alpha$ photons,
this ratio is unity. In young and compact planetary nebulae dust heating by direct absorption of starlight
is important, and therefore this ratio can be much larger than unity.
For this reason the infrared excess is often used as an age indicator for PNe.
Since the expected total infrared flux can be expressed in terms of optically thin radio flux density, we may
derive IRE from the formula given by Pottasch (\cite{Pottasch84a}).
In particular, as the nebulae associated to our targets  are compact, the infrared excess is derived in the
high density approximation, i.e.:
\begin{equation}
\mathrm{IRE}= 1.07 \, \frac{ F_\mathrm{IR}}{ F_\mathrm{8.4~GHz}}
\label{eq8}
\end{equation}

where $F_\mathrm{IR}$, in $10^{-14}\,{\rm W\,m}^{-2}$,  is the total infrared flux. For consistent comparison with other similar studies, $F_\mathrm{IR}$ has been derived by fitting a Planck curve  over the IRAS bands, namely from 12 to 100 $ \mathrm{\mu m}$ (Aaquist \& Kwok, \cite{AK90}).
Our sample is characterized by IRE values much higher than unity.
As a by-product of this method we also derive the dust color temperature.
A summary of the nebular characteristics  of our sample is reported in Table~\ref{detected}.

\section{Comparison with another sample}

\begin{figure}
\resizebox{\hsize}{!}{\includegraphics{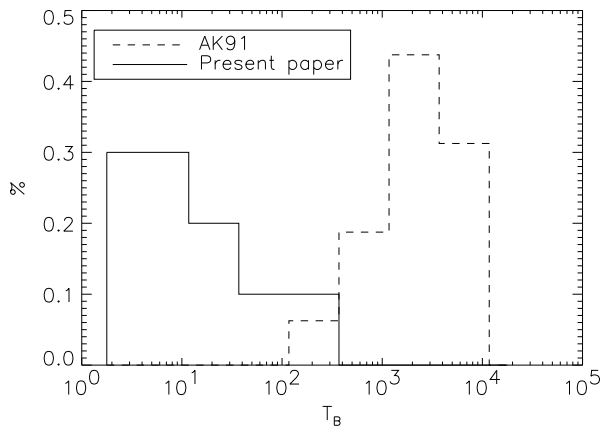}}
\caption{Distribution of brightness temperature ($T_\mathrm{B}$) for the AK91 sample (dashed line) and the
sample of the present paper (full line).
$T_\mathrm{B}$ for AK91 were calculated from Eq.(\ref{eq4}) and using radio measurements obtained at 15~GHz. Four of our targets have only lower limits for $T_\mathrm{B}$}
\label{tb}
\end{figure}
In order to check if our sample consists of young PNe, we will
compare the radio and infrared properties of the targets belonging to our sample
with results obtained in similar studies.
In particular, in the following, we will consider
the work of
Aaquist \& Kwok (\cite{aaquist}, AK91) carried out with the VLA at 15~GHz
on a sample of young PNe selected on the basis of their compact radio morphology.
The targets of AK91 were also observed at 5~GHz, but we prefer
to compare our results with those obtained at 15~GHz since  in both cases optical depth effects should
not be  important.\\
All the selected targets in AK91 have high brightness temperatures, infrared excess (IRE) much higher than
unity and dust temperature higher than the typical value observed in more evolved nebulae,
which is of the order of 100~K (Pottasch et al., \cite{Pottasch84b}).
All these properties are consistent with the hypothesis that  the sample  consists of very young PNe.

\begin{figure}
\resizebox{\hsize}{!}{\includegraphics{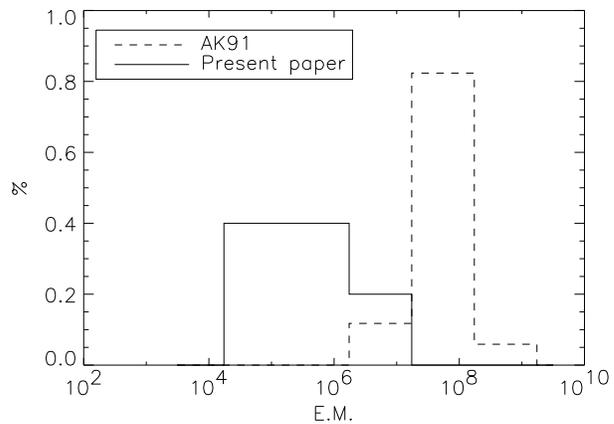}}
\caption{Distribution of mean emission measure (EM)  of the AK91 sample (dashed line) and of the present
sample
(full line).
Four of our targets have only lower limits for  EM }
\label{em}
\end{figure}

\begin{figure}
\resizebox{\hsize}{!}{\includegraphics{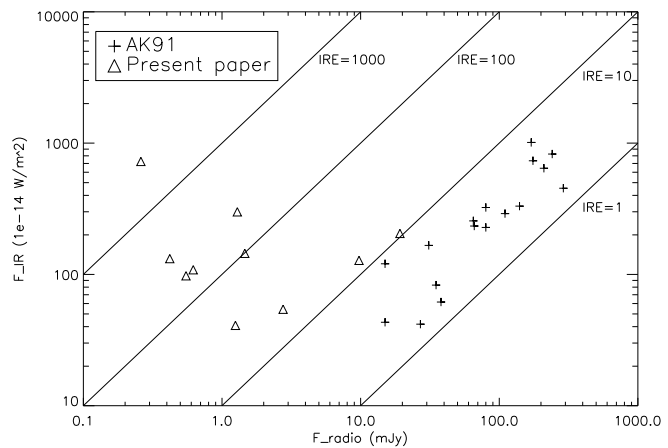}}
\caption{The radio flux density versus the far infrared flux for the present sample
(empty triangle) and for the AK91 sample (plus signs)}
\label{ire}
\end{figure}

In order to compare the physical properties of the nebulae belonging to AK91 with our sample we
plot in Figs.~\ref{tb},~\ref{em} and \ref{ire} the brightness temperature
($T_\mathrm{B}$), the emission measure (EM) and the infrared excess (IRE) of both samples.
Those quantities were re-calculated  from the published  radio measurements
using the same formulas as for our sample.

\begin{figure}
\resizebox{\hsize}{!}{\includegraphics{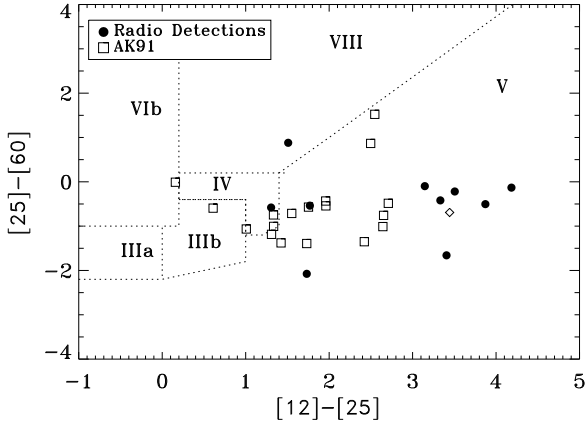}}
\caption{IR color-color diagram of detected sources (filled  dots) sources. Open squares are data from AK91.
The position in the color-color diagram of the very young
PN SAO~244567 is also marked as an empty diamond. Regions in the IRAS two-colors diagram for
different kinds
of source are also indicated (following van der Veen \& Habing, \cite{vandeV})}
\label{excess}
\end{figure}

It is evident that for the AK91 sample $T_\mathrm{B}$ and EM are systematically higher than for
our sample, and this seems to indicate  that our  sample  indeed consists of  more evolved PNe.\\
On the contrary, the infrared excess of our sample, which has values
systematically higher than those reported by AK91, indicates a particularly young sample of PNe.
This apparent contradiction is further complicated by the fact that the infrared properties
of both samples are quite similar, as is evident from an inspection of the IRAS color-color diagram
(Fig.~\ref{excess}), where sources  belonging to different samples  occupy the same region of the diagram.
This region is also shared with  SAO244567,  the youngest known PN,  whose evolution appears to be quite rapid,
since it has become ionized only within the past 20 years (Parthasarathy et al., \cite{Part93c}).\\

\begin{figure}
\resizebox{\hsize}{!}{\includegraphics{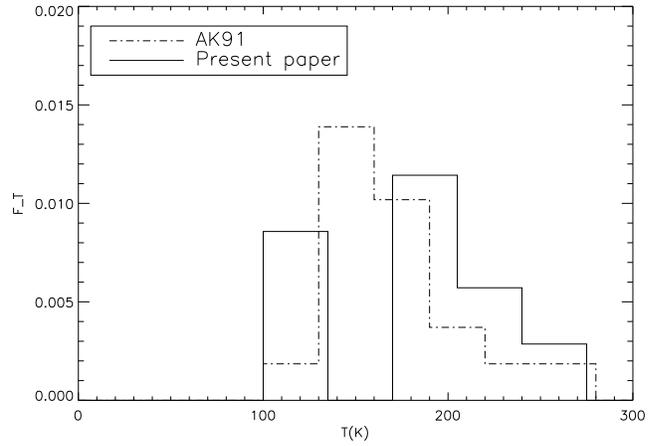}}
\caption{Distribution of dust  temperature of the AK91 sample (dashed line) and of the present sample
(full line)}
\label{dust}
\end{figure}

In the diagram the infrared colors are defined as usual (van der Veen \& Habing, \cite{vandeV}):
\begin{equation}
[12]-[25]=- 2.5\, \log{ \frac{F_{12}}{F_{25}}}
\label{eq9}
\end{equation}
where $F_{12}$ and $F_{25}$ are the fluxes measured at 12 and 25 $\mu$.\\
Moreover,  dust temperatures of both samples  have quite similar distributions (Fig.~\ref{dust}),
strengthening the conclusion  that both samples have similar dust characteristics.\\
Concerning the  possible causes of lower $T_\mathrm{B}$ and EM of our sample we may consider whether this could be
related to a systematic effect due to the different spatial
resolution used in the two surveys, as both $T_\mathrm{B}$ and EM depend on the angular source size
($\propto \theta^{-2}$).\\
A factor 20 in spatial resolution yields  a two orders of magnitude difference in both
$T_\mathrm{B}$ and EM,
forcing the two distributions to be more similar  than what is shown in Figs.~\ref{tb} and ~\ref{em}.
However, only 4 out of 10 detected sources were not resolved and thus without a  good angular size estimate.
Therefore, while this effect cannot be ruled out, at least for these 4 sources,
we will consider other possibilities which may explain our results.

\begin{figure}
\resizebox{\hsize}{!}{\includegraphics{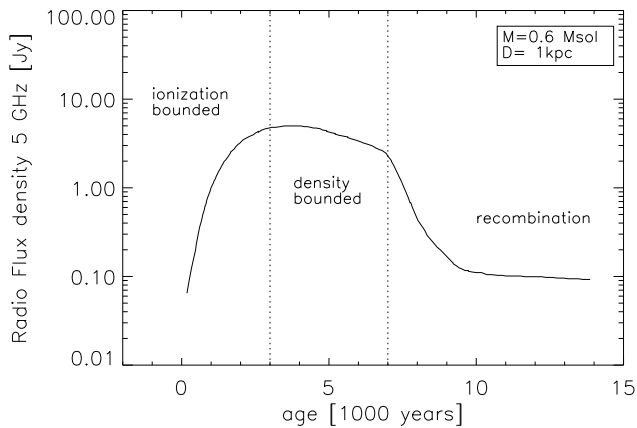}}
\caption{Evolution of the radio flux density in the framework of the dynamical model of
Volk \& Kwok (\cite{Volk}).
Adapted from Volk \& Kwok (\cite{Volk}). The curve is obtained assuming a nebula at a distance of 1 Kpc,
with a core-mass of
$ 0.6 M_{\sun}$ in the hypothesis of  a stellar wind with high  mass-loss
($10^{-4} M_{\sun}\,\mathrm{yr}^{-1} \leq \dot{M} \leq 10^{-5} M_{\sun}\,\mathrm{yr}^{-1} $)}
\label{evolution}
\end{figure}

Volk \&  Kwok (\cite{Volk}) studied the dynamical evolution of a nebula in the framework of the
interacting wind model.
They considered all the possible factors  that can affect the dynamics of the ejecta,
including radiative cooling,
variation of mass-loss of the stellar wind as well as  evolution of the central object.\\
The authors  also reproduced, in the context of their dynamical model, the time evolution of
the radio flux
density, assuming that all the radio emission arises only from the ionized part of the shell.\\
The evolution of the radio flux density with the age of the nebula is
outlined in Fig.~\ref{evolution}, adapted from Fig. 15 of Volk \& Kwok (\cite{Volk}), where
three distinct phases of radio emission have been indicated.\\
In particular, during the first phase, when the radio emitting region is bounded by the ionized front,
the radio flux increases with the radius of the ionized nebula, which is growing with the age of the nebula.
This is strictly related to the rate of Lyman continuum photons which increases with the central star
evolution
(Kwok, \cite{}1985).

In this context we may try to solve the apparent inconsistency between the typical values
of $T_\mathrm{B}$ and EM, as derived  for our sample,
which indicate a more evolved sample, and the IRE values, which, on the contrary,
appear to be consistent with a population
of very young PNe.\\
If we consider the functional dependence of $T_\mathrm{B}$, EM and IRE on the radio flux
density ($F_\mathrm{radio}$), the first two being proportional
to $F_\mathrm{radio}$  and  IRE to $F_\mathrm{radio}^{-1}$, our contradictory result
can be  explained if we assume that, on average, our sample is characterized by a very low radio flux
compared to what was
measured in the AK91 sample.\\
We thus  conclude that  the present sample is, indeed, very young and that the ionization  in the circumstellar nebula has just started.\\
Therefore, only a small fraction of the nebula is ionized and in the evolutionary scheme,
reproduced in Fig.~\ref{evolution}, they occupy the very initial portion of the
ionization bounded phase, corresponding to a very low value of radio flux density.
On the contrary, the AK91 sample, even if consisting of young PNe, occupies a more advanced
portion of the same evolutionary phase, corresponding to a much higher radio flux.

To further test our hypothesis, we calculate the ionized  mass of the nebula from the observed radio flux,
following Pottasch (\cite{Pottasch84a}):
\begin{equation}
M_{i}= 5.11 \times  10^{-5} (F_{8.4} \,\theta  \,d^{5})^{1/2}
\label{eq10}
\end{equation}
where it is assumed that the nebula has a filling factor $\epsilon=0.6$ and that $10\%$ of its mass  consists
of helium atoms.
In the formula $F_{8.4}$ is in mJy, $\theta$ is in arcsec, $d$ in kpc and $M_\mathrm{i}$ in $M_{\sun}$.
If we assume a standard distance of 1~kpc for the sources detected in our survey, we
obtain values for the ionized mass of the nebula  in the range
$6 \times 10^{-5} \leq \frac{M_\mathrm{i}}{M_{\sun}}\leq 5.4 \times 10^{-4}$,
which are much lower that the typical value of $ 0.2 M_{\sun}$
obtained for more evolved nebulae (Pottasch, \cite{Pottasch84a}).
If we assume that all the nebulae are at   a standard distance of 10 kpc, we obtain values
$2 \times 10^{-2} \leq \frac{M_{i}}{M_{\sun}}\leq 17 \times 10^{-2}$. These are to be considered
upper limits.

A radio nebula was recently detected around  SAO244567 (Umana et al., \cite{Umana2004}) and  preliminary
results were  presented by Trigilio et al. (\cite{Trigilio2003}).
A $T_\mathrm{B}=700 $  K  and a $EM=1.5
\times 10^{7} \mathrm{cm^{-6} pc}$ place SAO244567 between our sample and AK91 and this implies, in the framework of our hypothesis,
 that this young PN possesses a radio nebula slightly more evolved than those in our sample.

\section{Summary}

The formation and early evolution of Planetary Nebulae is an
unclear phase of stellar evolution.
It is still not well understood how the circumstellar envelopes, which form during the AGB phase,
transform themselves into the complex morphologies often observed in evolved PNe (Sahai, \cite{Sahai02};
Kwok et al., \cite{Kwok00}). In order to understand the origin of such  morphologies, it is  very important
to clarify at what stage of PN evolution the nebula becomes asymmetric or bipolar.
Moreover, recently  some objects have been identified whose evolution
towards the PN phase occurs on  timescales twenty times shorter than what is
foreseen by current evolutionary models (Parthasarathy, \cite{Part00}).
In this context it appears to be quite important to find objects in transition towards the PN phase,
which can provide strong constraints on determining  which process  shapes the nebula
and how this influences the nebular evolution.

In order to improve our knowledge in this field  we
have observed a sample of 16 hot post-AGB stars, selected on the basis of their optical
and infrared properties.
A common feature of our targets  is an infrared excess, signature of mass-loss that occurred during the AGB phase.
The main result of the present work is the detection of free-free radio emission in ten
of the observed sources. This indicates that ionization has already started in their
circumstellar material and that the sources can be considered to be in a  very early stage of Planetary
Nebulae evolution. Therefore, the detected sources constitute a unique sample  to be studied to
for shedding  light on this quite obscure evolutionary phase.\\

To further test our results, we  made a comparison with another sample of PNe assumed to be young.
We preferred the sample studied by Aaquist \& Kwok (\cite{aaquist}),
as both samples were observed at high frequencies and any possible opacity effect can be neglected.
By comparing the nebular characteristics of both samples, as derived from radio measurements,
we conclude that our sample comprises extremely young PNe.
The apparent   inconsistency between radio and infrared properties can be explained by assuming  that in our
nebulae the ionization has just started, while in the nebulae from AK91
the ionization structure is more evolved.
This translates into  a difference in the radio emission, which is, on average,
less intense for our sample.
This result is strengthened by the fact that we derive values of
ionized mass of the nebula which are much lower than those usually derived for more evolved nebulae.\\
Successive multi-frequency and high spatial resolution radio observations
 will allow us to fully characterize
 the radio properties of these new objects, providing important clues to better understand
 the formation and shaping of PN in the early stages of their evolution.

\begin{acknowledgements} This research has made use of the SIMBAD database, operated at CDS,
Strasbourg, France.
\end{acknowledgements}

\end{document}